# Trace Element Behavior during Shock Transformation of Zircon to Reidite

A. A. Shiryaev[a, *], A. N. Zhukov[b], V. V. Yakushev[b], A. A. Averin[a], V. O. Yapaskurt[c],

A. Yu. Borisova[d], A. Yu. Bychkov[c], O. G. Safonov[c, e, f], and I. V. Lomonosov[b]

a) Frumkin Institute of Physical Chemistry and Electrochemistry, Russian Academy of Sciences, Moscow, 119071 Russia

b) Federal Research Center of Problems of Chemical Physics and Medicinal Chemistry, Russian Academy of Sciences, Chernogolovka, Moscow oblast, 142432 Russia

c) Faculty of Geology, Moscow State University, Moscow, 119991 Russia

d) Géosciences Environement Toulouse, GET, Université de Toulouse, Toulouse, France

e) Institute of Experimental Mineralogy, Russian Academy of Sciences, Chernogolovka, Moscow oblast, 142432 Russia

f) Department of Geology, University of Johannesburg, Johannesburg, South Africa

*Corresponding author e-mail: a_shiryaev@mail.ru

## Abstract

Large single crystals of natural zircon were shock-loaded at 13.6 and 51.3 GPa in planar geometry. No structural changes were observed in zircon after loading at 13.6 GPa. Loading to 51.3 GPa resulted in zircon transformation to a denser scheelite-structured phase, reidite. The investigation of reidite samples by X-ray diffraction, Raman, photo- and cathodoluminescence spectroscopies revealed segregation of some trace cations (e.g., REE) on planar defects during the transformation. Importantly, the segregation occurred in a laboratory experiment without long-term annealing after shock loading. A possible mechanism of segregation of trivalent trace cations in zircon includes local violation of charge balance during the zircon–reidite reconstructive transformation, which is accompanied by considerable changes in the topology of polyhedra and second coordination spheres (Si–Zr). This results in expulsion of a fraction of the trace element into energetically expensive interstitial positions with high diffusivity even at relatively low temperatures.

**INTRODUCTION**

Zircon ($ZrSiO_4$) undergoes a structural transition at high pressure forming a scheelite-structured phase, reidite (Reid and Ringwood, 1969; Liu, 1979; Mashima et al., 1983; Kusaba et al., 1985). The transition pressure depends strongly on temperature and compression conditions: under static compression, the transition is observed at pressures higher than 19.7–23 GPa at room temperature, whereas the minimum pressure decreases to 8–10 GPa at 1100–1900 K. The transition is probably strongly affected by the kinetic factor (Gao et al., 2022). The existence of an intermediate phase was demonstrated at pressures higher than 10 GPa (Mihailova et al., 2019; Stangarone et al., 2019); it can transform back to zircon at decreasing pressure. Two mechanisms of the zircon–reidite transformation have been considered in the literature, diffusion-free martensitic and reconstructive. It was supposed that the martensitic mechanism is typical of crystalline zircon, whereas metamict domains transform reconstructively (Erickson et al., 2017). However, the detailed analysis of the zircon–reidite transition at static compression with the participation of the intermediate phase indicated that the reconstructive mechanism can occur also in crystalline material (Mihailova et al., 2019).

Reidite was documented in meteorites (e.g., Xing et al., 2020) and various products of impact metamorphism (Bohor et al., 1993; Glass and Liu, 2001; Wittmann et al., 2006; Chen et al., 2013; Zhao et al., 2021; Glazovskaya et al., 2024). It is important that it was found not only in relatively young impact structures but also in Precambrian rocks (Reddy et al., 2015; Li et al., 2018), which suggests that this high-pressure phase can survive over geological time. The presence of reidite and/or products of its transformation back to zircon constraints the minimum pressure that the rock experienced (e.g., Timms et al., 2017). Since zircon is extensively used in geochronological investigations, a question arises on the behavior of admixtures in this mineral during its complete or partial transformation to reidite. For instance, it was shown that the phase transition has no statistically significant effect on the U–Pb system (Szumila et al., 2023).

This paper reports the results of the investigation of the behavior of some trace elements during shock loading of a monocrystal of natural zircon to pressures of 13.6 and 51.3 GPa under laboratory conditions.

**EXPERIMENTAL AND ANALYTICAL TECHNIQUES**

In the shock compression experiments, 1.4-mm thick plates (Fig. 1a) cut from a natural zircon single crystal from the Mud Tank deposit in Australia (Gain et al., 2019) were used as starting samples in conservation capsules. The orientation of the plates was approximately (100). The samples were placed in 1.4-mm deep holes 20 mm in diameter in inserts in the capsules; the voids were filled with Wood's metal to minimize additional heating, material ejection, and other side effects (Fig. 1b). Then, the capsules were pressurized using a 50-ton hydraulic press.

The samples were shock-compressed in a close to [100] direction in planar geometry (Fig. 1c). The shock waves were generated by aluminum projectiles 5.0 and 7.0 mm thick accelerated by detonation products of an explosive or directly by an explosive planar shock wave generator to velocities of 3.3 and 1.15 km/s. The maximum pressures in the capsules were 51.3 and 13.6 GPa, respectively.

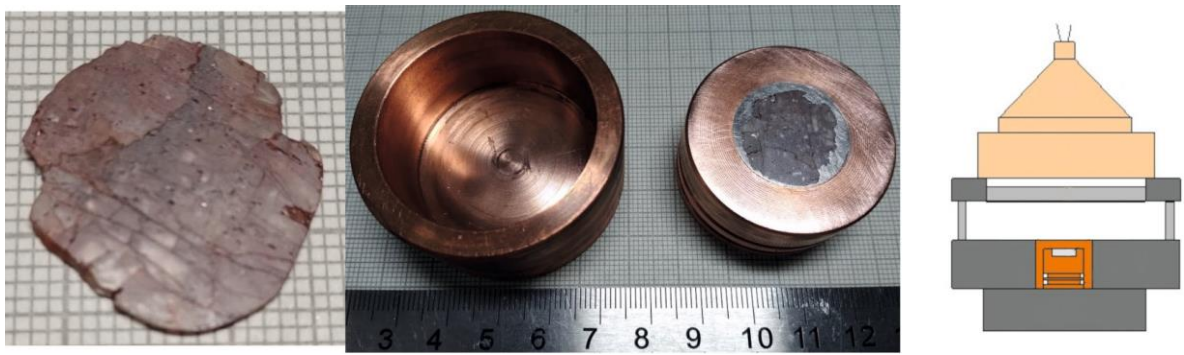

**Fig. 1.** Sample and details of experimental assembly: (a) plate cut from a zircon single crystal, (b) recovery ampoule with the initial sample, and (c) scheme of the experimental assembly: (1) detonator; (2) generator of a planar shock wave; (3) explosive pellet; (4) projectile; (5) steel ring; (6) acceleration flight path of the projectile; (7) copper recovery ampoule; (8) sample; (9) circular gaps; (10) protective steel ring; (11) steel base plate.

Figure 2 shows the calculated pressure profiles in zircon samples at shock compression in recovery capsules. They were obtained using a program for one-dimensional hydrodynamic calculation by the method of individual particles in cells (Fortov et al., 2006) and the equation of state of zircon derived using approach similar to that of Yakushev et al. (2019). The equation of state was based on isothermal compressibility data (van Westrenen et al., 2004) and the Grüneisen parameter at 300 K (Chiker et al., 2016). The agreement of such calculations with experiments was demonstrated previously by Yakushev et al. (2016). The maximum pressure was 13.6 and 51.3 GPa. As can be seen in Fig. 2, the sample pressure increased stepwise owing to multiple shock wave reflections from the walls of the copper capsule. The segment of constant parameters is followed by a pressure decrease to the atmospheric (null) value. The difference in the duration of high pressure influence in the profiles is related to the different thicknesses of the projectiles, which regulates, in particular, the launch velocity.

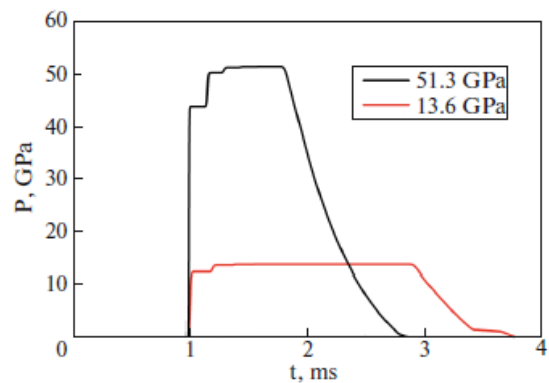

**Fig. 2.** Calculated pressure profiles in ZrSiO$_4$ samples during shock compression. See text for the calculation procedure.

Calculations above 30 GPa are complicated by the phase transition of zircon to reidite (Kusaba et al., 1985), the equation of state of which cannot be derived due to the absence of necessary data. In order to simplify the calculations for shock compression to 51.3 GPa, the phase transition was ignored, and it was assumed that only the low-pressure phase is compressed. Such an approach will presumably result in temperature overestimation owing to the additional splitting of the shock wave front related to the phase transition and a decrease in the duration of the maximum pressure in the profile before unloading compared with the experiment.

Temperature estimation using the equation of state showed that the samples were heated by 24°C during loading to 13.6 GPa and cooled by 3°C during unloading. Thus, the residual heating of the samples was 21°C. The heating, cooling, and residual heating during loading to 51.3 GPa were 265, 83, and 182°C, respectively.

Table 1. Results of the calculation of heating in the shock wave, cooling during unloading, and residual heating.

| $P$, GPa | | 13.6 | | | | 51.3 | | | |
|---|---|---|---|---|---|---|---|---|---|
| Porosity, % | | 0 | 1 | 5 | 10 | 0 | 1 | 5 | 10 |
| Heating in the shock wave, °C | Mixture | - | 47.2 | 124.5 | 196.7 | - | 344 | 664 | 1051 |
| | Medium | 24.5 | 23.6 | 21.3 | 20.0 | 265 | 256 | 230 | 201 |
| | Pore | - | 2359 | 1983 | 1610 | - | 8978 | 8466 | 7851 |
| Cooling during unloading, °C | Mixture | - | 5.70 | 15.2 | 23.8 | - | 110 | 220 | 350 |
| | Medium | 2.9 | 2.8 | 2.6 | 2.4 | 83 | 80 | 72 | 63 |
| | Pore | - | 291 | 243 | 195 | - | 3065 | 2876 | 2648 |
| Residual heating, °C | Mixture | - | 41.5 | 109.3 | 172.9 | - | 234 | 444 | 701 |
| | Medium | 21.6 | 20.8 | 18.7 | 17.6 | 182 | 176 | 158 | 138 |
| | Pore | - | 2068 | 1740 | 1415 | - | 5913 | 5590 | 5203 |

*Variations in temperature for the bulk porous sample are designated as "Mixture." Pore-free and strongly porous (50%) zircon in the sample are designated as "Medium" and "Pore," respectively. The assumption of the compression of the low-pressure phase only can result in the temperature overestimation (see text).

The optical microscopy of the starting samples revealed numerous fractures resulting in a decrease in mean sample density, which can be treated as the appearance of effective porosity. Therefore, we estimated the maximum sample temperatures for several porosity values. The temperature can be calculated for a porous sample as a whole (sample with averaged porosity) and separately for the material near pores and in unfractured areas. The temperature in material near pores was calculated considering the fracture and part of adjoining material as zircon with a high porosity of 50% and the remaining sample volume as a pore-free zircon. It was assumed that the pore is completely closed during shock wave loading, which is plausible for pressures above 10 GPa. The proportion of the massive and porous phases is controlled by the relative area of cracks, i.e., the porosity of the bulk sample. The maximum temperatures for pressures of 13.6 and 51.3 GPa were estimated made for sample porosity values of 0, 1, 5, and 10% (Table 1). The heating of material near pores is usually very high and increases with increasing shock pressure. Therefore, the temperature values near pores should be considered as approximate, since for the calculations experimental constants under standard conditions were used.

After the shock wave experiment, the capsules were turned and opened on a lathe. The approach used to the capsule opening preserves almost exactly the orientation of the samples during shock loading. After the opening, the remaining part of the copper capsule served as a substrate for X-ray diffraction investigation; to prevent material loss, it was impregnated with colophony solution in acetone. After the shock wave experiment at 13.6 GPa, the remains of the capsule lid were easily removed (Fig. 3). The removal of the lid and sample recovery after loading to 51.3 GPa required capsule heating with boiling water or the use of a dryer for Wood's metal melting and mechanical treatment (Fig. 3). The recovered samples were rinsed with 30% $HNO_3$ without heating for 20 min, washed in distilled water, and dried. After X-ray diffraction analysis (see below), the samples were cut perpendicular to the loading plane and mounted in epoxy for further investigation. Two polished sections were investigated for each of the two shock-loaded samples.

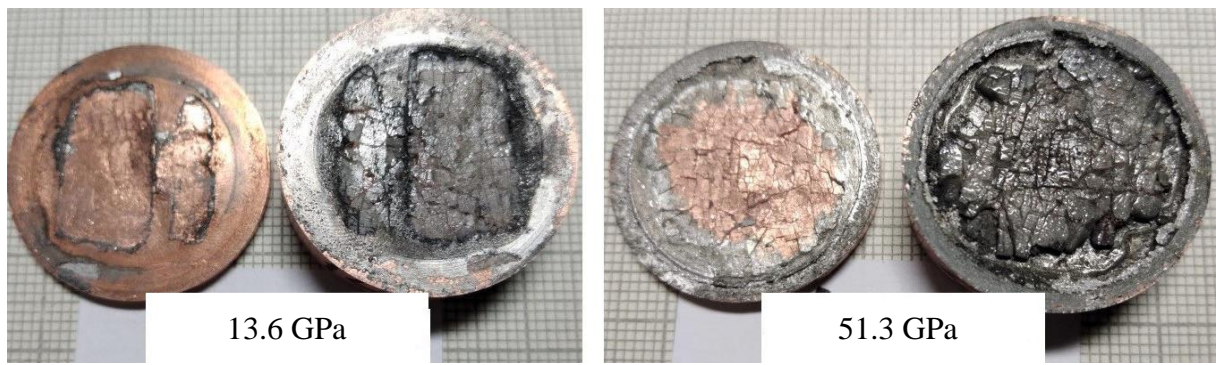

**Fig. 3.** Photos of the samples in the capsule after shock compression to 13.6 and 51.3 GPa.

The samples were investigated by X-ray diffraction, Raman photoluminescence and cathodoluminescence spectroscopy. The X-ray diffraction patterns were acquired directly from the surface of the sample attached to the copper substrate (see the description of the shock wave experiments) in Bragg–Brentano geometry using a DRON-4 diffractometer with

CuKα radiation. The spectroscopic investigations were conducted in spot and mapping modes. Color cathodoluminescence images were obtained without sample coating with a "cold" cathode at an accelerating voltage in the plasma cloud of 3–6 keV and a current of 100–300 μA. The spectra were recorded at ambient temperature using an Ocean Optics MayaPro 2000 spectrometer, contribution of the background luminescence of plasma was subtracted from the obtained spectra. The size of the area analyzed depended on the field of view, which was controlled by the objective used (5, 10, and 20×); the diameter of the analyzed area at the maximum magnification was ~200 μm.

Raman and photoluminescence maps were acquired by an inVia Reflex (Renishaw) spectrometer with 405 nm excitation. The luminescent lines were identified using published data (Krasnobaev et al., 1988; Nasdala et al., 2003; Friis et al., 2010). Infrared reflection spectra were recorded using a Nicolet iN10 (Thermo Scientific) IR microscope with a 50-μm aperture. Electron microscope images were obtained on a JEOL JSM IT-500 instrument equipped with BSE and SE detectors (Oxford).

## RESULTS

*Phase Composition and Orientation Relations*

The X-ray diffraction pattern of the starting crystal (Fig. 4) shows very strong 200 reflection of zircon and weaker 400, 600, and 800 reflections. In addition, there are weak 301 and 701 peaks. A narrow peak at 146° without characteristic splitting into the $\alpha_1\alpha_2$-doublet can be assigned to a reflection from zircon plane (901) for β-radiation. No foreign phases were revealed in the X-ray pattern. Thus, the system of the (200) planes of the zircon crystal is approximately parallel to its surface, although some individual grains have a different orientation.

In the impact experiment, the surface of the crystal plate was perpendicular to the direction of planar shock wave propagation. The high intensity of the (h00) reflections after loading to 13.6 GPa indicates pronounced zircon texturing. There are also weak peaks corresponding to the powdered zircon. Unidentified reflections are probably related to Wood's metal. In general, despite the sample fracturing clearly detected by visual examination, the fragments of the zircon crystal mostly retain their orientation, whereas the brecciated material filling fractures is disoriented. No evidence of polymorphic transformations was observed.

After shock compression to 51.3 GPa, the X-ray diffraction pattern changes completely, the peak positions correspond to reidite with the lattice parameters a = 4.722(2) Å, c = 10.524(5) Å, V = 234.7 Å3, and ρ = 5.189(8) g/cm3; there are also weak reflections of relict zircon. The extremely high intensity of the 112, 224, and 336 reflections suggests that the reidite grains are textured, and the (112) planes are oriented parallel to the sample surface. The obtained result (112)R||(200)Zrn supports the known zircon–reidite orientation relations (Leroux et al., 1999).

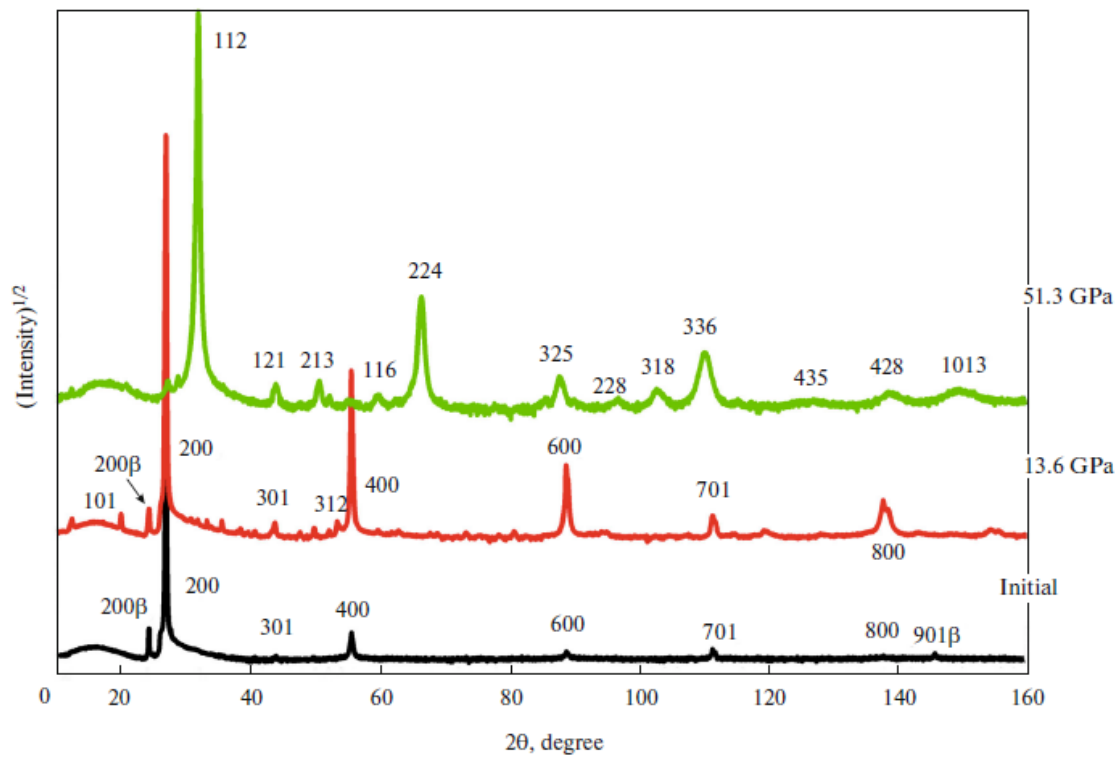

**Fig. 4.** X-ray diffraction patterns of the starting zircon and material after shock compression at 13.6 GPa and 51.3 GPa. The main reflections are labeled.

*Scanning Electron Microscopy*

Both shock-loaded samples are strongly fractured (Fig. 5). The brecciation of the material is stronger in the sample loaded to 13.6 GPa, and the size of unfractured areas is smaller. The contrast in back-scattered electron images is uniform (Fig. 5a). Loading to 51.3 GPa resulted in the appearance of numerous thin fractures subparallel to the shock wave plane and the (001) plane of zircon (Figs. 5b–5d). Perhaps, the developed microcleavage is related to the back reflection of waves during the experiment (analog of spall). Similar patterns of brittle fracture and dependence of the cracking on applied pressure were reported previously for both experimentally loaded natural zircon (Leroux et al., 1999) and natural samples from impact structures (Plan et al., 2021). The formation of larger blocks in the sample loaded to higher pressure was probably related to the fact that the brittle fracture began before the zircon–reidite transformation (Leroux et al., 1999).

The contrast of back-scattered electron images is very nonuniform at a scale of a few tens of micrometers (Fig. 5). However, this is not related to differences in chemical composition (for example, variations in Hf content) and is not manifested in cathodoluminescence images (see below). The lighter color in the back-scattered electron images is due to the higher density of reidite (by ~9%) compared with zircon; the darker contrast indicates a higher fraction of residual zircon. The size and shape of light and dark areas change randomly over the sample section, there are areas with patchy distribution of domains and areas with bands or wedges, sometimes discontinuous, of the darker material. Similar patterns were described in natural samples (Plan et al., 2021).

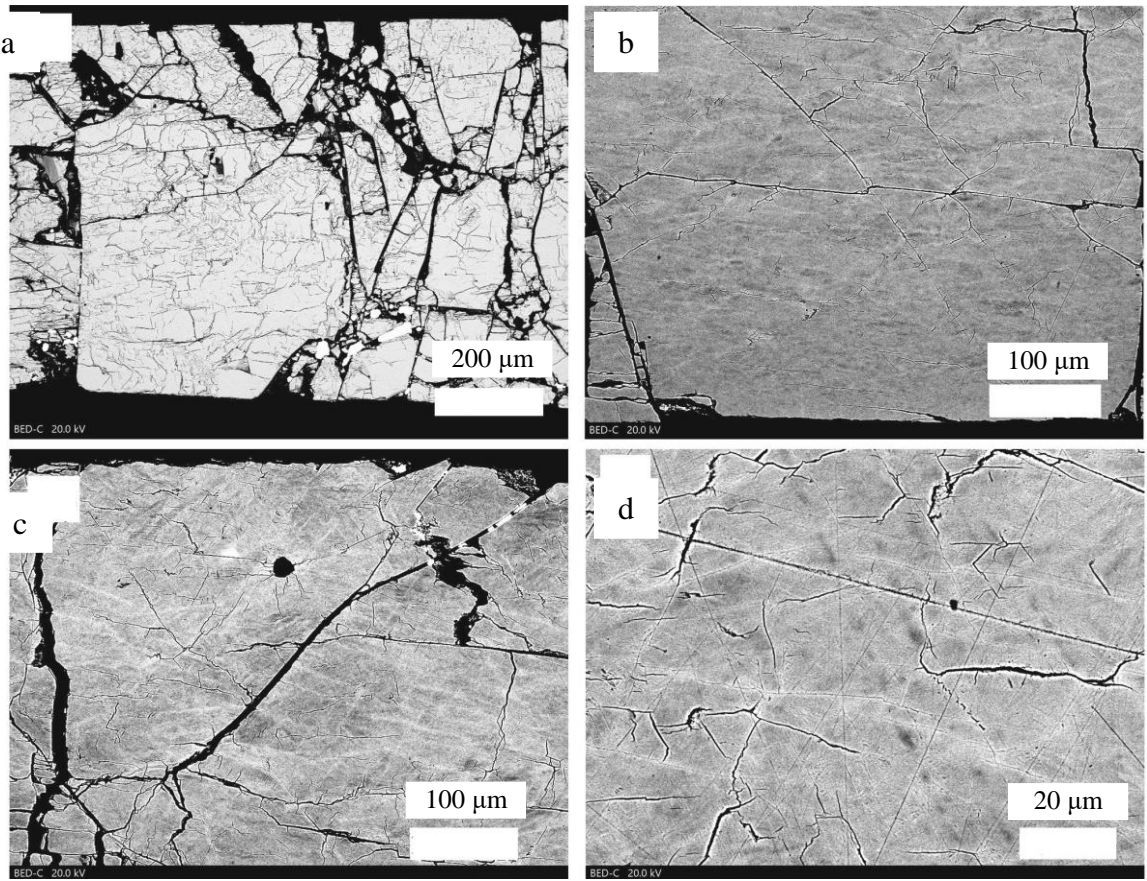

**Fig. 5.** Back-scattered electron images of samples (a) loaded to 13.6 GPa and (b–d) after the 51.3 GPa experiment.

*Raman Spectroscopy*

Numerous fractures appeared in zircon after loading to 13.6 GPa. The X-ray diffraction patterns and Raman spectra indicate the retention of the zircon phase, although the formation of reidite microdomains cannot be completely excluded. The microstructural features of this sample are discussed below during the description of cathodoluminescence investigations.

According to X-ray diffraction (Fig. 4), Raman spectroscopy (Fig. 6a), and IR spectroscopy (Fig. 6b), the material loaded to 51.3 GPa is represented by mosaic reidite. Figure 6a shows a representative Raman spectrum of this sample. Raman peaks were assignment follows Smirnov et al. (2010). The relative intensities of the reidite peaks $\nu_3z$ (845 cm$^{-1}$), $\nu_2xy$ (463 cm$^{-1}$), and $\nu_2z$ (401 cm$^{-1}$) vary considerably across the sample (see Fig. 4 of Gucsik et al., 2004) and indicate well developed mosaicism in the sample (see below). Although the spectra support the quantitative transformation to the reidite phase (Figs. 6c, 6d), weak peaks of residual zircon were observed at some spots (Fig. 6a). The most intense peak $\nu_3z$ of zircon is shifted relative to its position in the initial sample (1005 cm$^{-1}$) by 5 cm$^{-1}$ to lower values (1000 cm$^{-1}$) and is broadened at least six-folds (~17 cm$^{-1}$ versus ~3 cm$^{-1}$), indicative of high stresses in the lattice (see also Gucsik et al., 2004a). Although the spatial distribution of relict zircon is difficult to determine accurately due to the low intensity of the lines, mapping of the spatial distribution of the intensity ratios of the $\nu_3z$ line of zircon to all main peaks of reidite suggests that zircon is more abundant near fractures (Figs. 6c–6e). This observation can be explained by the lower peak pressure and more intense heating near fractures and/or in embrittled zones in the initial material.

*Photoluminescence and Cathodoluminescence*

The initial sample shows greenish cathodoluminescence (CL), the main volume of the sample is rather homogeneous, and there is no zoning (Fig. 7a). Variations in CL intensity are mainly explained by light scattering on internal fractures and other defects inclined relative to the polished sample surface. Nonetheless, large-scale mosaicism is manifested by a decrease in CL brightness in some parts of the crystal. The intensity of zircon CL excited by a focused electron beam in an electron microscope depends on the crystallographic orientation of the sample relative to the direction of the incident beam (Cesbron et al., 1995). However, in our study, plasma excitation of CL ("cold" cathode) was used, the exciting beam is defocused, and the orientation dependence of luminescence intensity is weak. The observed mosaicism can hardly be explained unequivocally because of the influence of a number of factors on CL intensity, including variations in chemical admixtures, perfection of the crystal lattice, and orientation of individual microblocks of the crystal (e.g., Rémond et al., 1995).

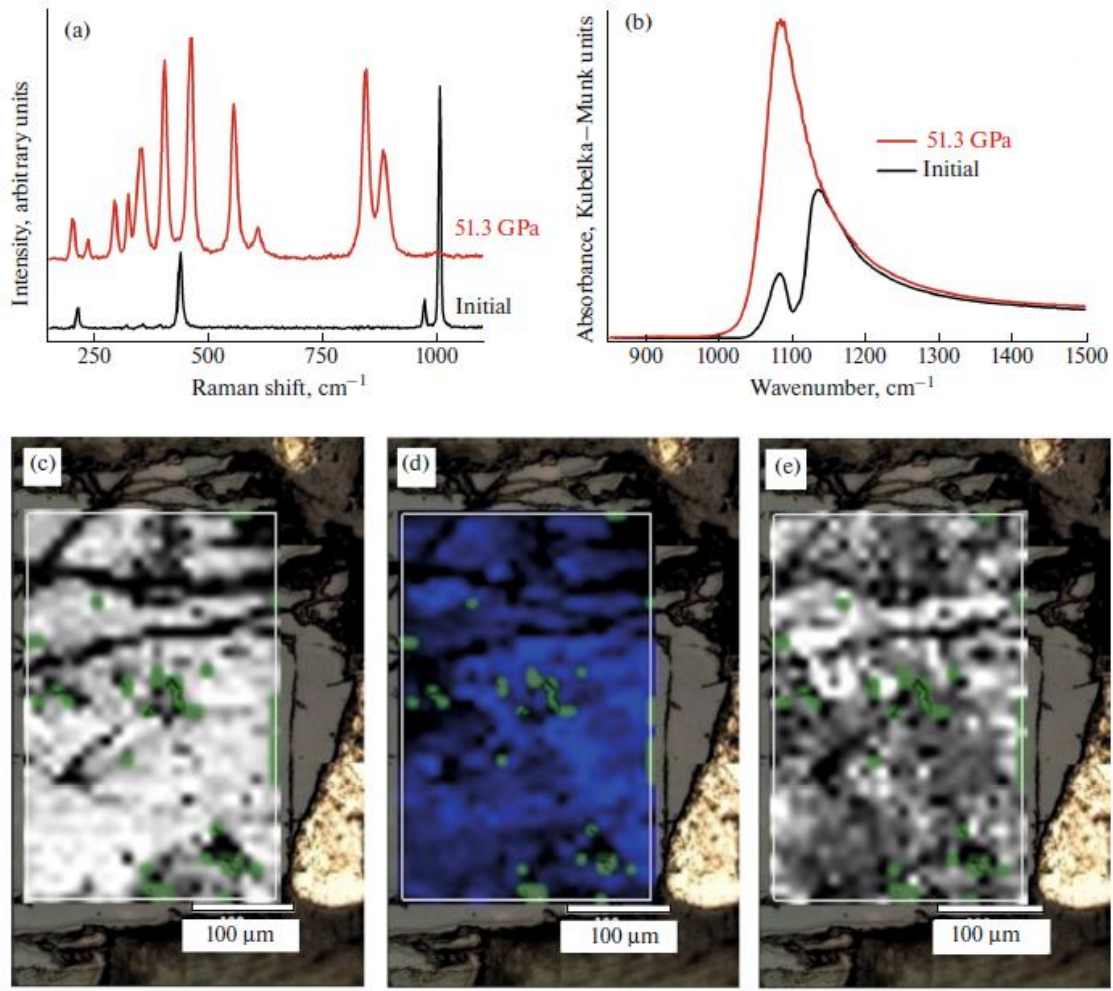

**Fig. 6.** Results of the spectroscopic investigation of the sample after loading to 51.3 GPa. (a) Raman spectra. The spectrum of the shock-loaded sample was acquired in the area where relict zircon was present. (b) Infrared reflection spectra. (c) Map of the distribution of the intensity of 882 + 884 cm$^{-1}$ doublet in the Raman spectrum of reidite. (d) Map of the distribution of the intensity of the 404 cm$^{-1}$ Raman peak of reidite. (e) Map of the distribution of the intensity of the main Raman peak of zircon at 1005 cm$^{-1}$. Light/blue areas in maps (c)–(d) correspond to higher intensities. Green spots are artefacts.

The CL spectrum of the initial sample can be deconvoluted to several components (Fig. 8). The most intense band has a maximum at ~550 nm, there are also a broad band at ~610 nm and a weak broad component with a maximum at ~690 nm. In addition to the broad bands, there are peaks related to the luminescence of REE$^{3+}$. The 487 nm peak and the poorly resolved 570 + 575 nm doublet correspond to transitions in $Dy^{3+}$ ($^4F_{9/2} \rightarrow$ $^6H_{15/2}$ and $^6H_{13/2}$, respectively). There is also a series of weak peaks at 390–430 nm, which are probably related to $Er^{3+}$.

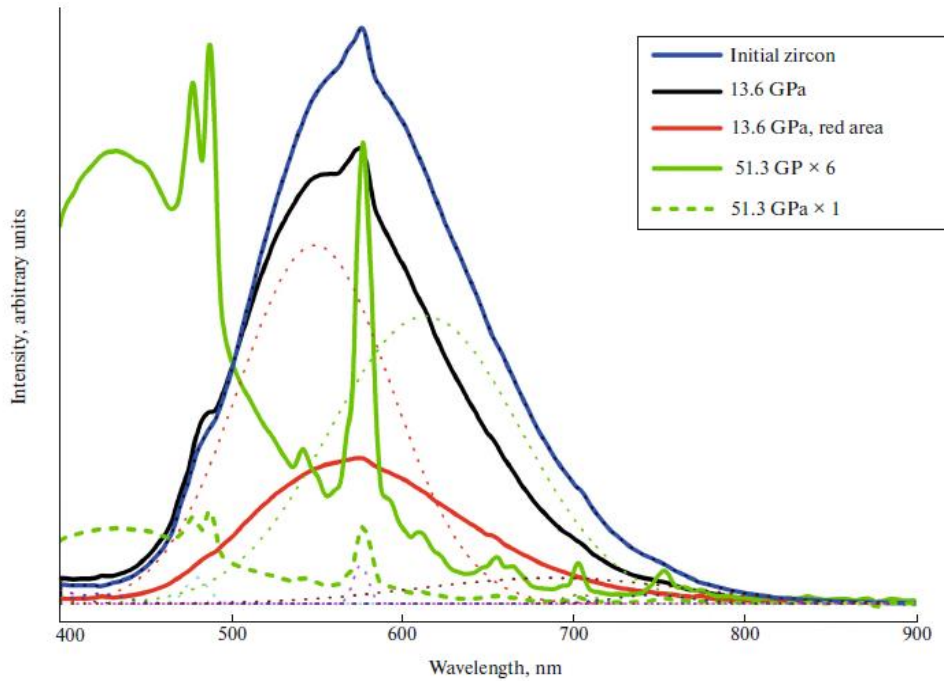

**Fig. 8.** Cathodoluminescence spectra of zircon before and after shock-wave experiments. The strong decrease in intensity for the 51.3 GPa sample in the short-wave region (less than 400 nm) could be related to absorption in the optical system (see text). The intensity of luminescence of this sample is shown both in real scale (thin dashed line) and after multiplication by six for better visualization. Thin dotted lines show main components of the starting zircon luminescence.

The CL spectrum of the sample loaded to 13.6 GPa is in general identical to that of the initial crystal (Fig. 8). However, the detailed inspection of images and adjustment of color balance revealed the presence of patches and diffuse areas, which possess a red tint and are sometimes adjacent to fractures (Figs, 7b–7d). The maximum intensity of luminescence in areas with a strong reddish tint is 3–4 times lower than that of the surrounding material. The intensities of the broad component with a maximum at ~600 nm and the 540 nm band are similar; the $Dy^{3+}$ peaks are very weak or absent. These two factors explain the reddish tint of CL and indicate the appearance of structural defects (e.g., vacancies and dislocations) favorable for radiationless transitions. Such areas could appear owing to an increase in the degree of sample mosaicism and, possibly, the higher degree of shock damage in crystallites with a certain orientation relative to the shock wave. One variant of such a scenario is partial transformation to the metastable "high pressure–low symmetry" ZrSiO$_4$ phase (Mihailova et al., 2019; Stangarone et al., 2019) and its retrograde transformation to zircon upon unloading accompanied by generation of point defects.

The CL of the sample transformed to reidite after loading to 51.3 GPa is approximately six times less intense than that of the starting material. The sample shows a blue luminescence with a maximum at ~430 nm; the relative intensity of CL at wavelengths <450 nm is higher than that of the initial material and the sample loaded to 13.6 GPa (Fig. 8). Since the optical system of the microscope and CL chamber absorbs radiation with wavelengths below 380 nm, it cannot be excluded that the luminescence band extends significantly into the UV region. Gucsik et al. (2002, 2004b) showed that the maximum of broad-band emission of experimentally loaded zircon lies at ~330–400 nm, with much weaker features above 500 nm. However, in these studies, much weaker spectral changes were produced by impact effects compared with the products of our experiments. The broad-band blue CL can be analogues to that observed in many silicates and related to charged SiO$^{4-}$ tetrahedra.

Along with the structureless band, there are narrow strong luminescence bands of REE$^{3+}$. The strongest 477 + 488 (clearly manifested doublet), 577 + 570 (poorly resolved doublet), 666, and 752 nm lines are related to Dy$^{3+}$ (transitions $^4F_{9/2} \rightarrow {}^6H_{15/2}$, $^6H_{13/2}$, $^6H_{11/2}$, and $^6H_{9/2}$, respectively). We also observed numerous weak bands probably related to Sm$^{3+}$ (568, ~605, 647, and 655 nm; transitions $^4G5/2 \rightarrow {}^6H_{5/2}$, $^6H_{7/2}$, and $^6H_{9/2}$), Pr$^{3+}$ (594 and 622 nm: $^1D_2 \rightarrow {}^3H_4$ and $^3F_0 \rightarrow {}^3H_6$), and Eu$^{3+}$ (611, 655, 693, and 702 nm; $^5D_0 \rightarrow {}^7F_2$, $^7F_3$, and $^7F_4$); some additional lines (542, 550, 560, 601, 622, 693, 743, 753, and 765 nm) could be related to Er$^{3+}$ and Tb$^{3+}$. A comparison of the CL spectra showed that after loading to 51.3 GPa intensity of REE$^{3+}$ lines increased relative to the broadband luminescence; some of the lines become narrower. The lines near 480 nm are related to the magnetic dipole transition $^4F_{9/2} \rightarrow {}^6H_{15/2}$ and are represented by well resolved doublet 477 + 488 nm. The ratio of the total areas of this transition and lines at 577 nm (electric dipole transition $^4F_{9/2} \rightarrow {}^6H_{13/2}$) remains almost unchanged or even decreases slightly. Note also that shock compression resulted in a considerable increase of the 577 nm component relative to the weak peak at 566–570 nm. These changes indicate appearance of at least two positions of Dy$^{3+}$ with different local symmetries in the reidite sample. Variations in the relative intensity and degree of splitting of the luminescence lines of Dy$^{3+}$ with increasing pressure in the shock wave and/or degree of impact metamorphism can also be observed in the CL spectra (see Figs. 15–17 of Gucsik et al., 2002).

An important difference of reidite (sample loaded to 51.3 GPa) from other samples studied can be seen in the CL images (Fig. 9): it contains numerous bright yellow speckles 2–10 μm in size (Figs. 9b, 9c), most of which are grouped. Many of the speckles appear to be confined to crystallographic planes inclined relative to the section surface. It cannot also be ruled out that the large structureless speckles are also confined to planes quasiparallel to the sample surface. The speckles are occasionally arranged in regular grids corresponding to the intersections of crystallographic planes (Figs. 9b, 9c). The angle between the grid lines is approximately 73°. Unfortunately, we failed to detect these features using scanning electron microscopy and photoluminescence mapping. These patches the specles are obviously related to the cathodoluminescence of $REE^{3+}$ and, first of all, the yellow band at 577 nm. Their origin is discussed below.

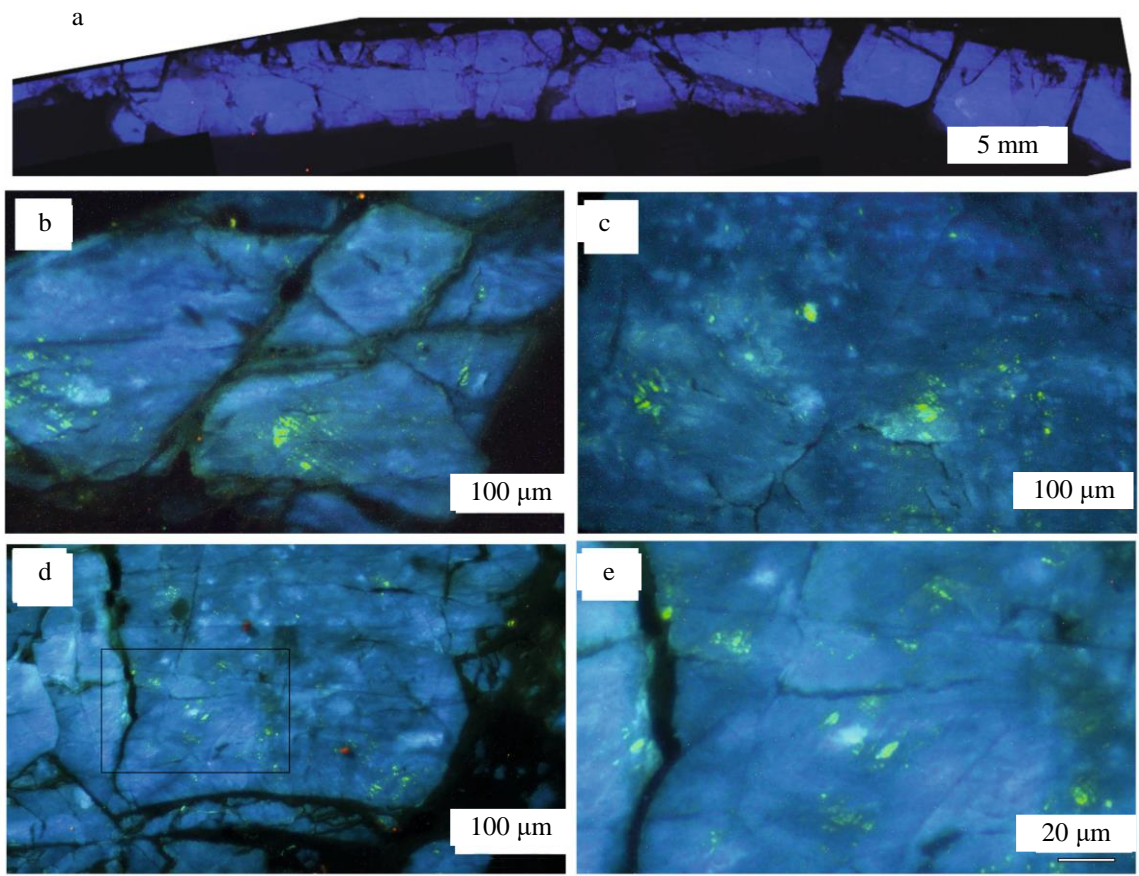

**Fig. 9.** Cathodoluminescence images of reidite sample after loading to 51.3 GPa. (a) Cross-section of the sample; shock wave propagated from top to bottom, i.e., in the [100] direction. (c) and (d) images of selected fragments of the sample after color adjustment, see text for explanation. The position of area (e) is shown by the rectangle in image (d).

The photoluminescence (PL) spectra of the samples excited by a 405-nm laser are significantly different from the CL spectra (Fig. 10). Broad bands are much weaker, and the number and intensity of REE$^{3+}$ peaks are highly variable. The higher resolution of the PL spectra is due to instrumental parameters and the significantly smaller excitation area (excitation laser beam is ~1 μm in diameter); correspondingly, the influence of broadening owing to mechanical stress and other similar factors is significantly weaker compared with the cold cathode CL spectra. The main features of the PL spectra are Dy$^{3+}$ (at 480–490 and 570–580 nm) and Eu$^{3+}$ (~610 nm) peaks; there are also peaks of Sm$^{3+}$, Pr$^{3+}$, and some other. The luminescence of Eu$^{3+}$ is more pronounced compared with the CL spectra. The relative intensity of the Eu$^{3+}$/Dy$^{3+}$ lines changed from point to point without any apparent correlation with other sample properties. The difference between the PL and CL spectra is probably related to the different excitation energies (3.06 eV at 405 nm and 3–4 keV, respectively) and corresponding mechanisms of energy transfer (e.g., Friis et al., 2010).

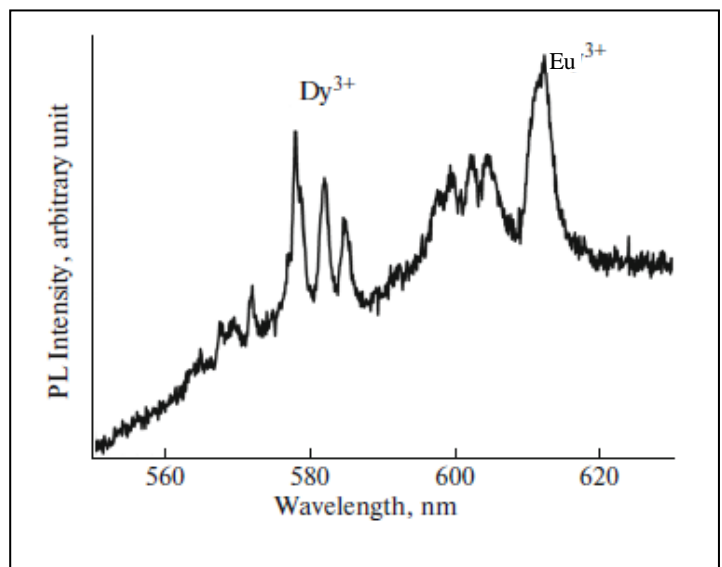

**Fig. 10.** Photoluminescence spectrum of reidite.

**DISCUSSION**

The investigation of natural zircons with reidite grains (Leroux et al., 1999; Montalvo et al., 2019; Szumila et al., 2023; Timms et al., 2017) and reidite obtained by laboratory dynamic loading (Gucsik et al., 2002, 2004b) revealed two main reidite morphologies: granular and lamellar. Granular reidite is probably produced from metamict zircon, and lamellae are characteristic of the transformation of a crystalline precursor (Erickson et al.,

2017). Plan et al. (2021) demonstrated the existence of three additional forms of reidite, including massive one characteristic probably of samples shock compressed to pressures higher than 50 GPa. The reidite sample obtained by us is probably a massive variety, although the spatial resolution of optical CL used by us can be insufficient for revealing fine lamellar structures. Note that the crystallographically ordered REE domains can indicate the presence of lamellar reidite.

The general intensity of CL decreases with increasing shock pressure, and the maximum of CL is shifted to shorter wavelengths (<450 nm) (Fig. 8). This observation is consistent with the published evidence of weak CL of reidite lamellae (Gucsik et al., 2002, 2004b; Plan et al., 2021). This is probably related to the appearance of numerous defects resulting in nonradiative losses. Changes in the luminescence lines of activator ions, including clearly manifested splitting of the lines of the magnetic dipole transition of $Dy^{3+}$ (477 and 488 nm) and significant changes in the relative intensity of the lines of electric dipole transition of $Dy^{3+}$, indicate the appearance of at least two positions of $REE^{3+}$ with different local symmetries in the reidite sample. It is logical to assume that these positions correspond to $REE^{3+}$ ions in the reidite structure and ions expelled to planar defects. The higher the symmetry of surroundings, the higher the relative intensity of lines related to the magnetic dipole transition (~480 nm) is. The bright yellow color of the segregations is related to the high intensity of the ~570 nm line, which is in good agreement with the low symmetry of the environment of $Dy^{3+}$.

It is known that in natural zircons that underwent impact transformation, trace elements, including Y, Yb, Al, Ca, Be, Mg, Mn, Ti, and Pb, are concentrated on low-angle boundaries in zircon and interphase zircon–reidite boundaries (Montalvo et al., 2019; Reddy et al., 2016). Segregation of trace elements on dislocation loops and planar defects of deformation origin was observed in zircon from metamorphic rocks and in experimentally annealed zircon (Peterman et al., 2019; Piazolo et al., 2016). One of the most intriguing features of the reidite sample synthesized by us is the appearance of speckles with enhanced $REE^{3+}$ luminescence. At least some of these domains are confined to planar defects and correspond very likely to segregation of luminescent ions on dislocations and planar defects generated during the impact transition of zircon to reidite. For instance, the shape of bright speckles in Figs. 9b–9d is qualitatively identical to the spatial distribution of trace elements documented by Montalvo et al. (2019, Fig. 4 therein). The aforementioned studies of Montalvo et al. (2019) and Reddy et al. (2016) did not reveal $REE^{3+}$ segregation, but this could be due to the poor detection limits of these elements by atom probe tomography because of the overlappings in mass spectra. Since the diffusion mobilities of $Dy^{3+}$, which is responsible for the bright luminescence of the segregations, and $Y^{3+}$ (Chernyak et al., 1997), are close it is reasonable to expect similar migration kinetics of these elements.

The segregation of REE ions on planar defects observed by us can probably be explained by a variant of the model of Montalvo et al. (2019), who considered mechanisms of trace element segregation on reidite lamellae in a zircon matrix from a large impact structure. However, this model requires substantial revision. Montalvo et al. (2019) assumed a diffusion mechanism for the trace element segregation during the cooling of shock-affected rocks over tens of days and more. In contrast to the previous studies, our samples underwent only minimum degree of annealing after the shock-wave compression. The segregation of trace elements in our experiments occurred on a time scale of tens of seconds, which indicates a significant role of the vacancy mechanism of diffusion and/or element transport along grain boundaries or dislocations.

The mechanism of the rapid trace element segregation remains elusive. The shock compression of materials is sometimes accompanied by extremely rapid formation of new phases and compounds, as well as diffusion of elements over macroscopic distances (e.g., Batsanov, 1996; Dremin and Breusov, 1968). In most cases, mass transport over macroscopic distances (millimeters) was observed in metals (e.g., Alekseevskii et al., 1989), in which various processes can be supposed, including turbulent mixing. However, there is an ample evidence of very rapid phase formation in dielectrics; this can be exemplified by $MgAl_2O_4$ spinel synthesis from individual oxides on phase boundaries (Potter and Ahrens, 1994), which was theoretically explained by Dremin and Breusov (1968) and Batzanov (1996). The mutual migration of K and Na ions through the boundaries of contacting feldspar grains over distances up to 300 μm was described in experiments on the shock loading of natural samples in converging shock waves (Kozlov et al., 1998). Moreover, the diffusion of these elements is controlled by the crystallography of mineral grains (Kozlov et al., 1998).

Trofimov et al. (2020) showed that the differences in the local surrounding of an admixture of group II elements in sphalerite at high and low temperatures are controlled, in particular, by the presence of charge compensating ions. For instance, gold admixture occurs as solid solution at high temperatures and forms AuS-like clusters during cooling (Filimonova et al., 2019). The retention of neutrality requires charge compensation accompanying REE$^{3+}$ substitution for Zr$^{4+}$ in zircon, which can be accomplished, for instance, by oxygen vacancies. It can be suggested that the transformation of the zircon structure to reidite via both the martensitic mechanism and, especially, reconstructive mechanism affects strongly the population of oxygen vacancies and similar defects. Changes in their local abundance can significantly reduce the probability of REE$^{3+}$ incorporation into the zircon structure and, thus, release part of them. If the expelled atom is, its high mobility even at moderate temperatures will undoubtedly promote segregation on extended defects. Note that natural zircon with reidite lamellae showed the maximum segregation of Al$^{3+}$, Ca$^{2+}$, Mn$^{2+}$, and Y$^{3+}$, the incorporation of which into the zircon structure requires charge compensation (Fig. 7b in Montalvo et al., 2019). It is possible that the diffusion of such expelled ions is responsible in part for the appearance of ionic and electronic conductivity of dielectrics and ionic crystals during the passage of shock waves (e.g., Al'tshuler, 1960).

In addition to the REE$^{3+}$ segregation on planar defects, other scenarios are possible. It is known that shock loading can be accompanied by partial zircon decomposition to oxides, ZrO$_2$ and SiO$_2$, and the submicrometer grains of ZrO$_2$ are confined to deformation lamellae in reidite (Reddy et al., 2015). Currently, we cannot completely rule out a connection between such features and the bright luminescent specles observed by us, but it is worth noting that neither baddeleyite nor orthorhombic ZrO$_2$ were observed in the Raman spectra and X-ray diffraction patterns. Part of REE ions can also entre the structure of relict zircon, but its volume fraction is minor.

**CONCLUSIONS**

The investigation of natural single crystal zircon shock compressed to pressures of 13.6 and 51.3 Gpa showed that the transformation to the high-pressure phase with a scheelite-like structure results in the segregation of some trace cations, for instance, REE, on planar defects. It is fundamentally important that such segregation was obtained under the conditions of a laboratory experiment, which did not include prolonged annealing of the material after shock wave loading. A possible reason for the segregation of substituting trivalent trace element in zircon at short time scales is the reconstructive mechanism of the zircon–reidite phase transition. Such a transition is accompanied by considerable reconstruction of the topology of polyhedra and second coordination spheres (Si–Zr pair), which may lead to a local charge imbalance and expelling of part of admixture atoms into interstitial positions. Since interstitial configurations are often energetically unfavorable, these atoms show high diffusion rates even at low temperatures.

The strong fracturing of shock-loaded zircon results obviously in an increase in material surface area. Taking into account the segregation of trace elements, their sudden release can be expected during crystal interaction with fluid or melt. Isotopic characteristics and trace element ratios in the lattices of zircon and reidite can remain approximately the same as in the starting zircon. It is interesting that the residual fracturing of zircon depends nonlinearly on the applied pressure: part of fractures could be closed after zircon transformation to reidite owing to material flow.